\documentclass[review,3p,times]{elsarticle}
\usepackage{lineno,hyperref}
\usepackage[T1]{fontenc}
\usepackage{ae,aecompl}
\usepackage{float}
\usepackage{balance}
\usepackage{mathrsfs}
\usepackage{enumitem, array}
\usepackage{amssymb}
\usepackage[labelfont=bf]{caption}
\usepackage{color}
\usepackage{multirow, booktabs}
\usepackage{tabularx}
\usepackage{amsmath}
\usepackage{comment}
\usepackage{hyperref}
\usepackage{bookmark}

\usepackage[tight-spacing=true]{siunitx}

\usepackage{calc}
\usepackage{graphicx}
\usepackage{pdflscape}
\usepackage{calc}

\usepackage{caption}
\usepackage{subcaption}
\usepackage[dvipsnames, svgnames, table]{xcolor}

\usepackage{parskip}

\hypersetup{
	colorlinks   = true, %Colours links instead of ugly boxes
	urlcolor     = red, %Colour for external hyperlinks
	linkcolor    = blue, %Colour of internal links
	citecolor   = blue %Colour of citations, could be ``red''
}

\journal{arXiv}

\begin{document}

	\begin{frontmatter}
		
		

		\title{How does surface wettability alter salt precipitation and growth dynamics during CO$_2$ injection into saline aquifers: A microfluidic analysis}
		
		
		\author[1]{Karol M. D\k{a}browski\corref{cor1}}
		\cortext[cor1]{Corresponding author}		
        \ead{karol.dabrowski@agh.edu.pl}
		\author[2]{Mohammad Nooraiepour}
		\author[2,3]{Mohammad Masoudi}
		\author[1]{Michał Zaj\k{a}c}	
		\author[1]{Szymon Kuczy\'nski}
		\author[1]{Rafał Smulski}
		\author[1]{Jan Barbacki}
		\author[2]{Helge Hellevang}
		\author[1]{Stanisław Nagy}				
		\address[1]{Faculty of Drilling, Oil and Gas, AGH University of Krakow, al. Mickiewicza 30, 30-059, Krakow, Poland}
		\address[2]{Department of Geosciences, University of Oslo, P.O. Box 1047 Blindern, 0316 Oslo, Norway}
        \address[3]{Applied Geoscience Department, SINTEF Industry, 7465 Trondheim, Norway}

		\begin{abstract}
        Carbon sequestration in deep saline aquifers is a promising strategy for reducing atmospheric CO$_2$ emissions. However, salt precipitation triggered by the evaporation of formation brine into injected supercritical CO$_2$ can cause injectivity and containment issues in near-wellbore regions. Predicting the distribution of precipitated salts and their impact on near-wellbore properties remains challenging. This study investigates the influence of surface wettability on CO$_2$-induced halite precipitation and growth within hydrophilic and hydrophobic microfluidic chips designed to mimic rock-structure porous geometries. A series of high-pressure brine-CO$_2$ flow experiments, direct microscopic observations, and detailed image processing were conducted to explore how substrate wettability affects salt precipitation. The experiments show that wettability markedly controls residual brine relocation, film flow movement, solute consumption, and salt formation. Tracking halite precipitation dynamics revealed distinct crystal formation signatures: hydrophilic chips exhibited irregular, larger aggregation patches, while the hydrophobic network showed more numerous, smaller, and limited aggregations. Large individual crystals were observed in both chips, with a notable dominance in the hydrophobic one. Crystallization dynamics varied, with nucleation and growth occurring earlier, progressing faster, and forming bulkier aggregates in the hydrophilic chip. Despite these differences, the three identified temporal stages of brine evaporation and halite surface coverage were comparable. Spatial analysis along the chips indicated that crystal aggregation properties, such as size and distribution, were position-dependent, with hydrophilic chips exhibiting greater probabilistic variability. These observations underscore the impact of surface wettability on salt precipitation through brine accessibility and capillarity, with implications for mitigating and remediating salt issues in saline aquifers.
		\end{abstract}
				
	
	\end{frontmatter}

\section*{Highlights}
\begin{itemize}
    \item CO$_2$-induced salt crystals precipitated in hydrophilic and hydrophobic porous chips.
    \item Dynamics, morphologies, distribution, and probabilistic nature were evaluated.
    \item Hydrophilic chip shows larger, irregular aggregation patches of micrometer-sized crystals.
    \item Earlier nucleation and faster growth dynamics observed in the water-wet chip.
    \item Large individual crystals dominate the hydrophobic chip.
\end{itemize}

\section{Introduction}
	\label{Introduction}
Carbon Capture, Utilization, and Storage (CCUS) offers an effective strategy to reduce atmospheric CO$_2$ emissions, a primary driver of climate change \cite{IPCC_2022_WGIII, ringrose2021storage}. CCUS technology encompasses capturing CO$_2$ from industrial point sources or directly from the atmosphere (DAC), utilizing it in various industrial processes, and ultimately storing it in geological formations. Depleted gas and oil reservoirs provide a means to sequester CO$_2$ securely, leveraging existing infrastructure and delivering enhanced hydrocarbon recovery. Saline aquifers, with their widespread availability and substantial storage potential, are particularly promising for large-scale sequestration projects. Additionally, CO$_2$ can be employed as a cushion gas in underground storage facilities. Further potential geological repositories include deep coal beds, where CO$_2$ can be adsorbed onto coal surfaces displacing methane, and mafic and ultramafic rocks where CO$_2$ can undergo carbon mineralization to form carbonate minerals \cite{oelkers2023carbon, gislason2014carbon, sandalow2021carbon}. 

Among the various geological carbon storage (GCS) candidates, deep saline aquifers are particularly favored due to their extensive availability, large storage capacity, and proximity to emission sources \cite{ringrose2021storage, krevor2023subsurface, Hellevang2015}. The effectiveness of CO$_2$ storage in any geological repository primarily depends on three criteria: injectivity, capacity, and containment. This paper focuses on injectivity, which is the ease with which CO$_2$ can be introduced and flow into the storage reservoir. The results also have implications for containment due to the potential for CO$_2$-induced salt crystallization damage in reservoir rocks \cite{nooraiepour2024Damage, shahidzadeh2024crystallization}.

The well-reservoir interface of a CO$_2$ injection well serves as critical gateways to the geological storage units \cite{ringrose2009plume, zettlitzer2010re, Khosravi2023}. Pore space in this area can become blocked due to various thermo-chemical processes such as mineral precipitation \cite{torsaeter2018geological, masoudi2021thesis, XIAO2024ReviewRisk}. One of the main geochemical phenomena in the near-wellbore reservoir zone, which has severe hydraulic and mechanical consequences, is the formation dry-out and salt precipitation.

Injecting dry CO$_2$ into underground reservoirs leads to evaporation of the formation brine. As the CO$_2$ injection continues, the brine in the vicinity of the injection wellbore starts to evaporate, increasing the concentration of dissolved salts in the remaining brine. When the salt concentration in the remaining brine exceeds its solubility limit, salt crystals begin to form and precipitate within the reservoir rock pore space \cite{nooraiepour2018effect, cui2023review, miri2016review, derluyn2024experimental}. This phenomenon, known as CO$_2$-induced salt precipitation, can significantly alter the permeability and porosity of the rock, leading to various operational challenges \cite{nooraiepour2018salt}. CO$_2$-induced salt precipitation has been studied and documented across various scales, from pore-level to field-scale \cite{talman2020salt, sokama2023experimental, miri2016review, cui2023review}, utilizing field observations, laboratory experiments, and numerical simulations \cite{mim2023minireview, MIRI201510,nooraiepour2018effect, ott2015salt, ott2021salt, NOROUZI2021CO2Salt, MASOUDI2021103475, HoTsai2020, Falcon-Suarez2020, LIMA2021FracDry, ZEIDOUNI2009600}. Numerous parameters influence drying in porous media, including the salt type, thermodynamic conditions, characteristics of porous media, and the interplay of various forces and transport phenomena \cite{Shokri2024soilSal, derluyn2024experimental}. These factors affect the dynamics of salt precipitation. However, many aspects remain unexplored, necessitating further research to fully delineate the multifaceted underlying physics.

Microfluidics is a particularly popular tool for studying salt precipitation because it allows for the visual investigation of processes at smaller scales \cite{Datta2023LabCarbon, Morai2020HPMicrofluidic}. This method enables to directly capture the growth behavior of salt crystals within the porous medium, providing valuable insights into the mechanisms at play \cite{HE2022ExpSalt, HoTsai2020}. Despite its potential, studies exploring the effect of wettability on solid precipitation through microfluidics remain sparse \cite{dTzachMineralWett, DARKWAHOWUSU2024Review}. Wettability describes the tendency of a fluid to spread on a solid surface in the presence of other immiscible fluids. Understanding wettability is crucial because it significantly affects fluid distribution in the porous medium, thereby influencing distribution and precipitation patterns.

Akindipe et al. (2022) \cite{akindipe2022salt} used a miniature core flooding setup under x-ray tomography and found that wettability significantly influences salt precipitation. In intermediate-wet carbonates, salt precipitation was initiated earlier and caused more porosity reduction than in oil-wet systems. Zhang et al. (2024) \cite{zhang2024brine} used air injections into low-pressure microfluidics to show that hydrophobic systems had less backflow and lower brine saturation, resulting in minimal salt accumulation. Yan et al. (2024) \cite{yan2024dynamics} found that intermediate-wet porous media exhibited reduced rates of both water evaporation and salt precipitation, while water-wet surfaces formed larger salt crystals.

He et al. (2019) \cite{he2019pore} found that wettability affects the location and structure of salt precipitation. On hydrophilic and neutral surfaces, salt precipitated ex-situ, clogging pore throats and reducing permeability. Conversely, hydrophobic surfaces showed in situ precipitation, causing less severe permeability loss. They also showed that increasing the CO$_2$ injection rate can suppress capillary backflow and minimize permeability reduction due to salt precipitation. Rufai and Crawshaw (2018) \cite{rufai2018effect} used microchips with varying wettability and showed that oil-wet systems exhibited a significantly shorter early stage of evaporation. In water-wet and mixed-wet systems, salt deposition more severely impeded permeability due to continuous hydraulic connectivity driven by capillary forces.

Wettability has been shown to significantly impact water evaporation and drying mechanisms. In hydrophobic media, evaporation rates are substantially reduced due to the absence of a continuous liquid film, which inhibits efficient water transport. Conversely, hydrophilic conditions facilitate faster evaporation owing to the film flow that promotes more efficient liquid transport \cite{shahidzadeh2004effect,bachmann2001isothermal,shahidzadeh2007effect,SHOKRI2008Hydrophob}. Changes in drying dynamics will, in turn, affect precipitation patterns \cite{bergstad2016evaporation,SHOKRI2013135,Desarnaud2015salt, PRAT2024}. While the effect of wettability on evaporation and drying dynamics has been studied in various contexts, few studies have specifically focused on its impact on CO$_2$-induced salt precipitation relevant to subsurface thermodynamic conditions. Wettability may also play a crucial role in managing potential adverse effects related to pore clogging in the near-wellbore region.

Another less investigated aspect of CO$_2$-induced salt precipitation is the probabilistic nucleation and stochastic growth of mineral formations in pore spaces. To capture natural reality accurately, it is essential to predict not only the amount but also the location and distribution of crystal nucleation and precipitation over space and time \cite{Nooraiepour2021SciRep, Masoudi20249988, Nooraiepour2021Omega}. Traditional reactive transport models, based on transition state theory (TST), classical nucleation theory (CNT), and most non-equilibrium surface models, fall short of answering the “where” question, encompassing location and distribution \cite{nooraiepour2022GHGT16}. 

To address this gap, we introduced the Probabilistic Nucleation Model \cite{Fazeli2020nucleation, Nooraiepour2021SciRep}. This model allows for studying the uncertainty related to porosity-permeability changes during subsurface mineral precipitation \cite{Masoudi20249988}, crucial for carbon and energy storage applications such as CO$_2$-induced salt precipitation in saline aquifers or carbon mineralization in basalt (mafic and ultramafic rocks) \cite{DENG2022105445, MASOUDI2021103475, Hellevang2019Basalt, EAGE2024Basalt}. By incorporating the intrinsic stochastic variations of natural geological media (heterogeneity and anisotropy), the probabilistic nucleation model can provide more precise, physics-based predictions for precipitation-induced pore clogging across scales \cite{Masoudi20249988, MASOUDI2021103475}, as one of the most uncertain aspects of the near-wellbore GCS simulations \cite{parvin2020continuum, MASOUDI2021103475, masoudi2021thesis}. Additionally, the probabilistic model allows for better simulation of coupled chemical-hydraulic-mechanical processes, which influence the fate of subsurface fluid flow and solute transport.

Building on the current understanding of wettability's impact on salt precipitation, this study aims to delve deeper into the pore-scale probabilistic dynamics of CO$_2$-brine transport and halite precipitation in porous microfluidic chips under high-pressure conditions. By examining hydrophilic and hydrophobic surface wettability, we seek to provide insights into how parameters such as brine transport, solute availability, film flow, and capillarity control the dynamics, morphology, and distribution of halite nucleation and growth within the pore network. This study will elucidate how these mechanisms contribute to the formation of large salt aggregation patches and their subsequent impact on porosity-permeability decline, flow obstruction, and potential pore clogging. By monitoring multiple cycles of evaporation and precipitation, we aim to analyze the spatial and temporal dynamics of crystallization. Additionally, through repeated experiments, we seek to quantify the probabilistic nature of halite nucleation and crystal growth, thereby understanding the variability in salt precipitation dynamics during CO$_2$-induced reactive transport processes.

	\section{Materials and Methods}
    \subsection{Microfluidic chips and laboratory setup}
    This study employs two porous media micromodels with rock-like geometries featuring hydrophilic and hydrophobic surface wettability (Micronit Micro Technologies). Each chip represents a realistic reservoir rock within a 20 mm (length) $\times$ 10 mm (width) domain. Within the porous domain, pore sizes range from 100 $\mu$m to 600 $\mu$m, with a median value of 250 $\mu$m. Additional properties of these microfluidic chips can be found in Table~\ref{tab:Table1}.
             
    \begin{table}[htp]
    \centering
	\caption{ \bf{Detailed properties of the hydrophilic and hydrophobic microfluidic chips.} }
	\footnotesize	
		\begin{tabular}{>{}m{1.8in}  >{\arraybackslash}m{1.8in}}
			\hline
			porosity  &   $0.47\pm1$\\
            permeability  & $7.2\pm1.1$  $D$ \\
            contact angle hydrophilic chip &  $27^\circ $\\
            contact angle hydrophobic chip &  $105^\circ $\\
   		chip thickness &	1800 $\mu m$ \\
			material & borosilicate glass \\
			\hline
		\end{tabular}
	\label{tab:Table1}
    \end{table} 

\begin{figure}
	\centering	
    \includegraphics[width=0.9\textwidth]{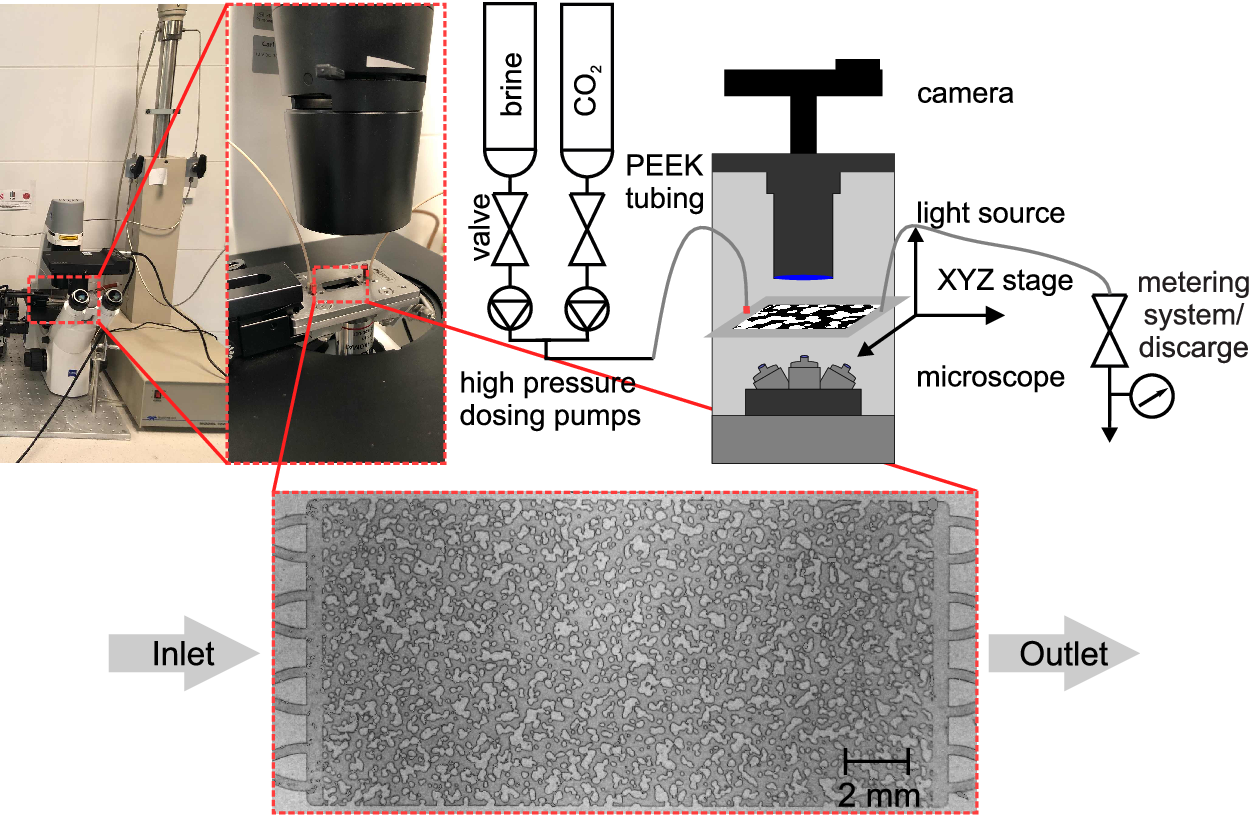}
	\caption{Schematic and photograph of the developed experimental setup. The microfluidic chip is mounted on an XYZ motorized stage, and images are recorded using a FASTEC IL5 high-speed microscopic camera. High-precision syringe pumps controlled brine and CO$_2$ injection. Flexible PEEK tubing interconnects the chip with the injection system, allowing for smooth movement during the experiments. Insert shows a microscopic image of the microfluidic chip used in the experiment. Arrows indicate the direction of CO$_2$ flow, displacing brine out of the pore space (injected from left to right) at a 2 ml/min flow rate at $22^\circ C$ temperature and 5 MPa pressure.}
	\label{fig:figure1}
	\end{figure}

    A custom-designed experimental setup was assembled for this study, as shown in Figure ~\ref{fig:figure1}. The microfluidic chip is held securely within a holder under a microscope, positioned on a light source, and connected to the fluid injection system via 0.25 mm flow tubes. The chip is mounted on an XYZ motorized stage (Standa Ltd.) for precise positioning during measurements. Flow-through experiments were monitored using high-speed microscopy to capture fluid movement and crystal growth. Spatial and temporal dynamics were tracked with a ZEISS Primo Vert microscope, which includes a binocular phototube, a HAL 35 W light source, and a 3 W LED. The microscope is equipped with a Plan-Achromat 4x/0.10 objective. Imaging was carried out using a FASTEC IL5 high-speed camera with a 2560 x 2080 12-bit CMOS sensor, 5 $\mu$m low-noise pixel size, and a frame rate of 20,000 fps.
    
    The fluid injection system used high-precision syringe pumps (Teledyne ISCO 30D) to control the displacement of brine and CO$_2$ at specified flow rates and pressures, while a backpressure regulator maintains internal fluid pressure.
    
    Insert in the Figure~\ref{fig:figure1} presents a microscopic image of the entire porous domain of the microfluidic chip, also highlighting the fluid delivery and distribution channels at the inlet and outlet. These channels distribute the injected fluid across the entire cross-section of the chip. Both the inlet and outlet are equipped with single $1/16''$ PEEK tubing. The microchips are fabricated from borosilicate glass and fused silica materials using an acid chemical etching method, which produces isotropic etching features and semi-homogeneous surface-edge conditions. The internal structure of the chips is randomly generated, resulting in slight variations in pore geometry between the two chips while maintaining statistically similar and representative reservoir rock geometries. 

    \subsection{Experimental procedure}
     We investigated the pore-scale dynamics of CO$_2$-brine transport and halite precipitation when dry CO$_2$ (unsaturated with respect to water) was injected into porous microfluidic chips. Initially, the porous network was saturated with a 5 M sodium chloride (NaCl) solution in deionized (DI) water. CO$_2$ was injected at an injection pressure of 5 MPa, a controlled flow rate of 2 ml/min, and a temperature of \(21 \pm 1\,^\circ\mathrm{C}\). During these tests, a backpressure regulator maintained the internal pore pressure at 2 MPa, and the effluent was collected at ambient conditions at the outlet.

    First, We conducted four experimental cycles using each hydrophilic and hydrophobic chip. Evaporation and crystal growth cycles were monitored over a fixed 2 × 2 mm area located 10 mm from the chip inlet. After each experiment, the chips were cleaned with DI-water and vacuum-dried. The cleaned chips were then placed in a vacuum chamber and fully saturated with the brine stock solution before each run. In subsequent experiments, the XYZ stage was adjusted during CO$_2$ injection to observe crystallization at various positions along the chips, reflecting different channel geometries and spatial evaporation-precipitation dynamics from inlet to outlet.

    \subsection{Digital image processing}
    To accurately track and quantify the displacement of fluid phases, evaporation of residual brine, and halite precipitation, we conducted a series of post-image processing steps on the captured snapshots to distinguish the involved phases. The recorded images were processed using a threshold algorithm to identify the boundaries of the porous structure, as well as the spatial and temporal evolution of the residual brine, CO$_2$ stream, and halite crystals. We adjusted the brightness to a desirable range to minimize background light deviation between images.
		
\begin{figure}
		\centering	
		\includegraphics[width=0.8\textwidth]{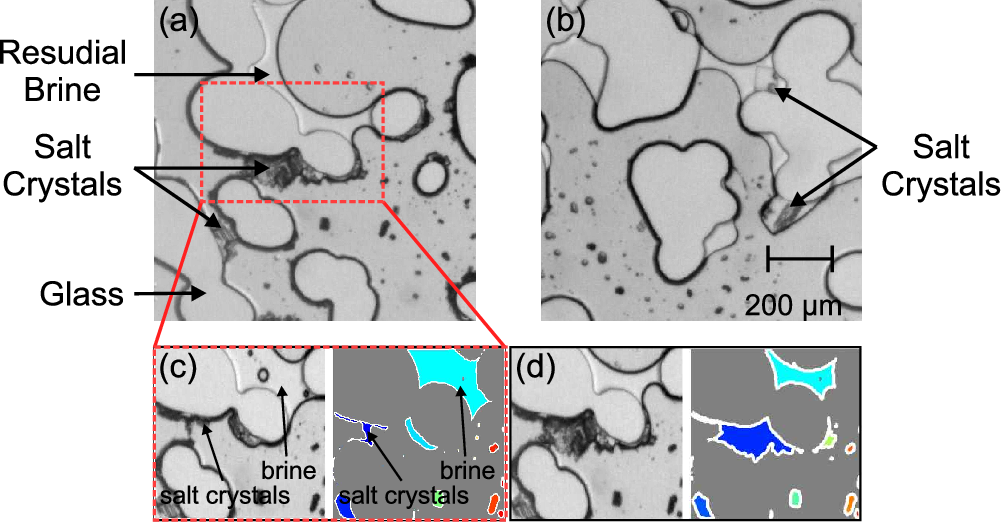}
		\caption{Microscopic view of (a) hydrophilic and (b) hydrophobic chips after partial brine dry-out. Arrows mark the locations of salt crystals, residual brine, glass, and pore space. As CO$_2$ was continuously injected, the salt concentration gradually exceeded the saturation level due to water evaporation, leading to salt precipitation. (c, d) Two distinct temporal stages of evaporation and precipitation within the hydrophilic chip. The left panels show microscopic visualizations of the flow experiments at high magnification, whereas the right panels display the detected crystalline structures and residual brine. }
		\label{fig:figure2}
	\end{figure}

    After binarizing and segmenting the raw images, three phases (CO$_2$, salt, and brine) could be visualized and quantified. We generated two binary sequences from each experimental series: (i) the drying process of in-situ brine saturation, and (ii) salt precipitation, growth, and aggregation. Figure~\ref{fig:figure2} (c,d) shows an enlarged area of the hydrophilic chip at two temporal stages of brine evaporation and crystal growth, along with the detected phases.
    
    The developed image analysis pipeline allows the determination of brine saturation $S_w = A_B/A$ and crystal areal coverage $X_c = A_C/A$ for each experimental image, where $A$ is the chip total area. The size distribution of individual crystals can be compared by calculating the equivalent diameter: $d_{\text{eqv},i} = \sqrt{\frac{4 \cdot A_{C,i}}{\pi}}$ and the diameter center of mass (CoM) defined as: $D_{\text{CoM}} = \frac{\sum_{i} N_{i} d_{\text{eqv},i}}{N}$, where $N$ is the total number of detected crystals. We calculated time-evolving features by measuring the coverage areas in these two binary image sequences over time ($t$). We computed the time derivatives of $dA_B(t)/dt$ and $dA_C(t)/dt$ to track the rates of brine drying and salt crystal precipitation, where $A_B$ represents the brine areal coverage and $A_C$ represents the halite precipitation area. By comparing the behavior of a given region over time, the algorithm determines the phase evolution. This procedure was performed for each region. 

    \section{Results and Discussion}
    \subsection{Dynamics of brine evaporation and halite precipitation}

    Upon CO$_2$ injection, brine is initially displaced via viscous two-phase flow. As the process continues, water molecules in the residual brine are carried out of the porous network by the CO$_2$ stream and evaporate, controlled by the solubility of brine into the CO$_2$ under the experimental thermodynamic conditions. This eventually leads to nucleation and precipitation of salt crystals as halite supersaturation is achieved and evaporation progresses within the chips.

\begin{figure}
		\centering	
		\includegraphics[width=0.9\textwidth]{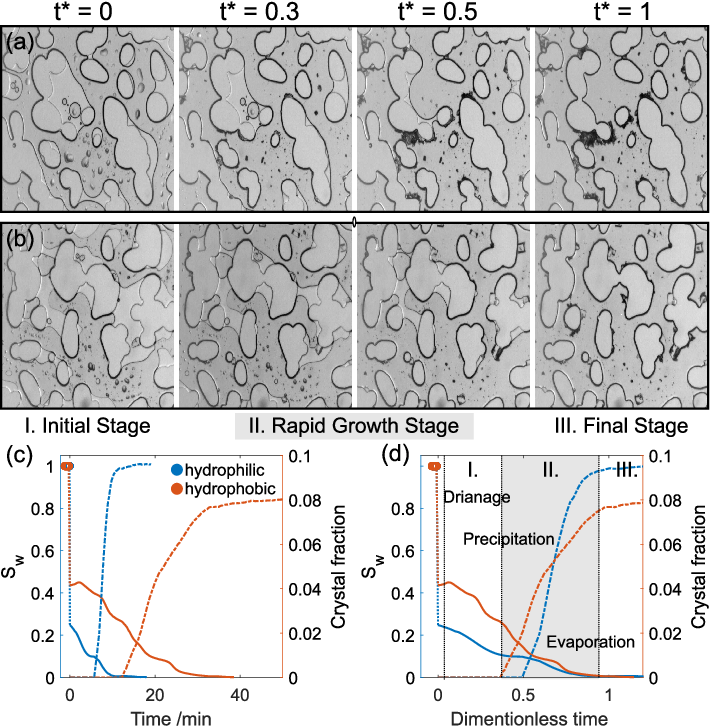}
		\caption{Temporal dynamics of brine saturation and salt precipitation as a function of elapsed and normalized experimental time. The experimental images show snapshots of porous networks after initial brine displacement, rapid evaporation, and salt crystallization stage, and the final phase of brine dry-out for (a / top) the hydrophilic chip and (b / bottom) the hydrophobic chip. (c,d) Brine saturation change (solid lines) and crystal growth (dashed lines) over time. Blue lines denote the hydrophilic chip, while red lines denote the hydrophobic chip.}
		\label{fig:figure3}
    \end{figure}

    The CO$_2$-brine interface shows distinct meniscus configurations and contact areas within the two chips due to their differences in wettability, as illustrated in Figures~\ref{fig:figure2} and~\ref{fig:figure3}. In the hydrophilic chip (Fig.~\ref{fig:figure3}(a)), water tends to form larger clusters in localized regions compared to the hydrophobic chip. Additionally, the hydrophilic chip features fewer flow channels and smaller CO$_2$ finger formations, indicating enhanced connectivity of the wetting phase within the pore network influenced by surface wettability. On a hydrophilic surface, the CO$_2$-brine menisci at the pore scale exhibit a pronounced water affinity. Consequently, the propensity for water to achieve long-range continuity through corner flow or thin-film flow is significantly higher than on a hydrophobic surface. Figure \ref{fig:figure2} (a) and (b) show enlarged areas of a single pore channel exhibiting partial crystallization within the hydrophilic and hydrophobic chips, respectively. Arrows indicate the locations of matrix grains (glass), precipitated salt crystals, and the remaining aqueous phase. In the hydrophilic chip, visible salt aggregation is shown where they block a local flow channel. The hydrophobic chip shows the halite precipitation with smaller and more numerous crystals.     
    
    Figure~\ref{fig:figure3} shows the overall temporal dynamics of brine saturation reduction and halite growth under viscous displacement, subsequent evaporation, and salt precipitation for hydrophilic and hydrophobic chips due to continuous CO$_2$ injection. It presents the computed brine saturation ($S_w$) and crystal areal coverage ($X_c$) (Figs.~\ref{fig:figure3} (c-d)) as a function of elapsed and normalized experimental time. The presented microscopic images capture a 2×2 mm area located 10 mm from the chip inlet. The total porosity of the chips at this specific position was 0.47. The first images of each row show both chips after initial brine displacement under drainage. The next two snapshots at $t^* = 0.3$ and $0.5$ illustrate further temporal stages of brine evaporation and crystal growth, where $t^*$ is the normalized time. The last snapshot of each row depicts the chips after complete dry-out.

    Evaporation is traditionally divided into different stages \cite{lehmann2008characteristic, or2013advances, bacchin2021microfluidic}, with subtrends resembling those shown in Figure~\ref{fig:figure3} \cite{HoTsai2020, rufai2018effect}. The early phase, characterized by a relatively high and constant evaporation rate, is known as stage-1 evaporation or the constant rate period (CRP). This phase is followed by a much lower evaporation rate, referred to as the falling rate period (FRP) and the receding front period (RFP). Often, the FRP and RFP are combined and collectively referred to as stage-2 evaporation. The transition from CRP to FRP occurs when gravity and viscous forces overcome the capillary driving force in porous geometries, severing hydraulic connectivity. Consequently, in our flow-through experiments, the CRP will end quickly due to the viscous displacement of the drainage stage, and the majority of observations occur under the FRP and RFP periods. This rapid transition shifts from an advection-dominated period to a diffusion-dominated period (RFP), accompanied by a further and prolonged reduction in the evaporation rate.
 
    In the rock-like pore geometries of both chips, an uneven distribution of diverse residual brine patches across the pore space becomes evident after the drainage stage. The heterogeneous pore structure of these reservoir-rock representative chips has the potential to retain more brine as residual saturation due to varying local capillary pressures. From the sequential snapshots and quantitative measurements in Figure~\ref{fig:figure3}, we observe three stages of brine saturation reduction following the initial full saturation state: (I) viscous displacement or drainage, (II) fast-progressing dynamics of drying of residual brine via evaporation into the CO$_2$ stream, and (III) end-stage slow decline until complete dry-out.

    These processes, occurring over different timescales, are evidenced by the initial rapid decline in $S_w$, subsequent slower changes in brine saturation, and gradual salt formation (shown by the appearance and accumulation of black spots between approximately 40-100\% of drying time in Fig.\ref{fig:figure3} (a,b)). Through quantitative image analysis of the time-varying coverage area, we measured the brine-drainage rate and subsequent evaporation rate to elucidate the corresponding dynamics and mechanisms at different temporal stages.    

    In the initial instances, the saturation of the pore structure is rapidly swept out as the gas injection pressure overcomes the capillary pressure, displacing the brine from the pores. The steep decline in the volume of residual brine illustrates this drainage phase. Here, the dimensionless time ($t^*$) is defined as $t^* = \frac{(t - t_0)}{(t_f - t_0)}$, where $t_0$ is the initial time of drainage (start of injection), and $t_f$ is the complete drying time.

    The post-drainage brine saturation ($S_w$) was calculated to be 0.248 for the hydrophilic chip and 0.436 for the hydrophobic chip, indicating higher residual brine in the non-water-wet pore space. After drainage, the residual aqueous phase becomes nearly immobile, trapped in specific pore areas, and forms various pools of various sizes, in addition to covering parts of the grain matrix. Because the injected CO$_2$ was unsaturated with water vapor (the solubility of CO$_2$ in water at the test conditions is 1305 $mol/m^3$ \cite{MOSAVAT20145408}, while the water content in the utilized scientific-grade gaseous phase was infinitesimal), water molecules evaporate from these brine patches into the gas stream. The rapid evaporation phase (stage II) exhibits three sub-trends of drying rates, with differing evaporation trends observed between hydrophilic and hydrophobic cases.

    As shown in Figure~\ref{fig:figure3}, the dynamics of halite precipitation were also computed and illustrated by the increasing area of salt crystals over time. Three stages of CO$_2$-induced salt precipitation were identified based on the observed growth rates: (I) initial salting-out, (II) rapid growth and aggregation, and (III) final gentle halite expansion. These distinct salt precipitation stages were consistently observed across repeated experiments with both chips. The period of faster evaporation dynamics corresponds to a more rapid stage of salt precipitation.

    The image analysis results in Figure~\ref{fig:figure3} indicate that the salt precipitation rate (unit pore size per $t^*$) during the initial stage (I) is close to zero. During the rapid growth stage (II), the halite growth rate accelerates to 0.29 and 0.13 (unit pore size per $t^*$) for the hydrophilic and hydrophobic chips, respectively. In the final stage (III), the salt growth rate slows down to 0.014 and 0.012. The drying rate of residual brine is 0.28 and 0.41 for the hydrophilic and hydrophobic chips during stages (I+II), and in the third stage (III), the drying rate of brine decreases to 0.002 and 0.026 before complete dry-out.

    The correlation between the halite growth rate and the drying/evaporation rate during the three stages mentioned above can be described as follows. In the initial phase, the residual brine is confined to the pore throats and narrower sections of the porous medium. Consequently, the evaporation rate remains low due to the limited contact area between the gas stream and the receding/evaporating aqueous phase. At this early stage, salt nucleation and precipitation occur slightly in the small residual brine areas and at the edges of brine pools adjacent to the main CO$_2$ flow pathways. Given the limited size of the brine contact area and the need to bring progressively more trapped brine to a supersaturated state, halite formation advances extremely slowly.

    Initially, the water will be undersaturated with respect to salt, and it will take some time to reach saturation and begin precipitation. Over time, diffusion will establish a more uniform and higher salinity throughout the water, leading to faster precipitation in the later stages. As more brine evaporates, the solute concentration within large brine pools or interconnected fluid patches gradually reaches the solubility limit, initiating rapid growth. This is due to the intrinsic fast nucleation and growth rate of halite crystals, the availability of a supersaturated aqueous phase, and an adequate gas-liquid contact area for evaporation within the porous medium. This rapid process is observed in the stage II of evaporation-precipitation. Two mechanisms likely induce this significant increase in the evaporation/growth rates. First, the gas-liquid contact front gradually shifts from the throats toward regions with larger pore spaces, enlarging the evaporation surface area, accelerating brine evaporation, and salt formation. Second, small salt crystals nucleated during the initial stage, which are not yet fully visible, help accelerate the evaporation and growth processes by creating a positive feedback loop as they form on or adjacent to previously precipitated salt crystals (secondary substrate) \cite{Nooraiepour2021Omega, Nooraiepour2023PorMedAlb, desarnaud2015drying}. Previously precipitated sites act as sweet spots for further nucleation and halite growth due to solute availability, brine continuity, and favorable surface free energy. These locations ensure a sustained supply of dissolved ions \cite{masoudi2024EAGE, nooraiepour2023EAGE} and provide conditions that lower the energy barrier for additional crystal formation \cite{Masoudi2022EAGE, nooraiepour2022GHGT16}, creating a self-reinforcing cycle we previously coined as 'salt self-enhancing growth' \cite{MIRI201510, miri2016review}.

    The hydrophilic nature of halite crystals nurtures this feedback loop by attracting surrounding brine to form a thin film on the crystal surfaces. This brine film and the often wet surfaces of halite crystals at this stage significantly increase the CO$_2$–brine contact area, thereby enhancing water evaporation. Concurrently, a new layer of salt crystals nucleates on the previously precipitated crystal surfaces after the water molecules evaporate from the brine film, extending the salt precipitation coverage \cite{nooraiepour2023EAGE, masoudi2024EAGE}. The newly formed salt creates more brine-film-covered areas, further promoting the positive feedback loop and enhancing additional nucleation. Large brine pools adjacent to the drying and nucleation/precipitation interfaces serve as primary ion sources during this rapid growth stage, continually providing sufficient solute to the precipitation region and maintaining the growth as capillary forces supply the solute transport. Consequently, residual brine is rapidly consumed, and extensive salt bodies precipitate, primarily as micrometer-sized porous and semi-porous aggregates.

    In stage III, as large brine pools dry out and the majority of halite solutes are depleted, the growth rate significantly slows. This slowdown occurs due to the difficulty in accessing fresh brine sources and the reduced salt concentration in the remaining brine, which fails to maintain a supersaturated state. Therefore, regenerating the supersaturation condition and nucleating the remaining salt ions takes longer, delaying the complete dry-out of the system.
		
    \subsection{Morphology of pore-scale halite precipitation}
\begin{figure}
    \centering	
    \includegraphics[width=0.9\textwidth]{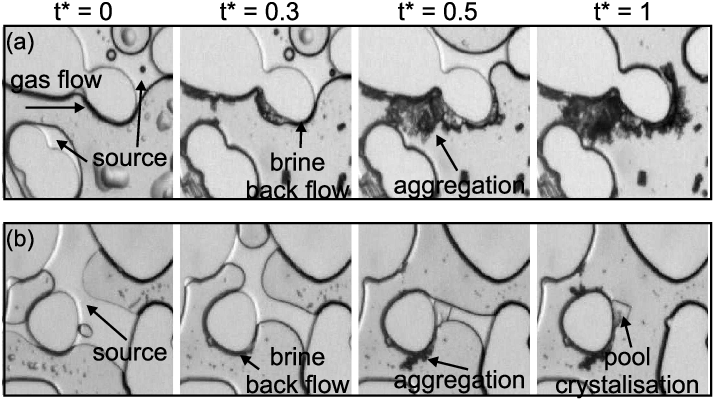}
    \caption{Comparative analysis of salt precipitation morphologies and growth patterns across different temporal stages and surface wettabilities. Each sequence of images details the progression of salt formation, beginning with the initial residual brine regions, followed by subsequent salt precipitation patterns. Microscopic photographs capture the temporal dynamics of brine drainage and salt precipitation, showcasing differences in behavior between (a) hydrophilic chips, which exhibit strong aggregation, and (b) hydrophobic chips, which display weak aggregation and pool crystallization.}
    \label{fig:figure4}
    \end{figure}

    Consistent with our previous research \cite{nooraiepour2018effect, nooraiepour2018salt, MIRI201510}, we observed two primary halite precipitation morphologies on the chips with different surface wettability, as shown in Figure~\ref{fig:figure4}: large individual halite crystals with fully developed crystalline shapes and polycrystalline aggregates of micrometer-sized halite (nano- to micron-sized crystals). The formations of individual crystals and polycrystalline aggregates are categorized into three halite precipitation stages, illustrated in Figure~\ref{fig:figure4}. Each row of image sequences demonstrates the evolution of salt growth dynamics before and after the appearance of the precipitates.

    At the initial precipitation stage for both chips, most salt crystals nucleate within the brine pools trapped in pore regions where viscous displacement could not efficiently remove the aqueous phase. These locations are often outside the main fast-flowing pathways. As these brine spots shrink due to drying, halite begins to nucleate on an interface (either the water-gas interface or the chip-water interface) and grow, consuming the residual brine. Large crystals, characterized by their semi-transparent and ordered cubic structure in Figure~\ref{fig:figure4}, nucleate through the self-assembly of solutes in the liquid phase. This occurs when the ion concentration exceeds the solubility limit under the given thermodynamic conditions. This process is driven by local evaporation rates that facilitate the necessary supersaturation for crystal growth. Eventually, the crystals extend into the surrounding gas stream, protruding into the gas-liquid interface \cite{nooraiepour2018effect}. These crystals form at moderate rates within the brine pools, developing their crystalline structures. 

    After the formation of these individual large crystals, the acceleration of evaporation and brine consumption rates leads to the formation of micrometer-sized halite aggregates. These aggregates appear as darker salt patches of various sizes at the interfaces with the CO$_2$ stream in Figure~\ref{fig:figure4}. Provided they receive a sufficient solute supply from adjacent brine sources \cite{masoudi2024EAGE}, these aggregates extend into disorderly precipitated salt crystals. A key feature of these aggregates is their intercrystalline porous structure (secondary porous medium) formed within the original geometries of reservoir rocks \cite{nooraiepour2023EAGE}. Significant differences between the two wettability cases become apparent during the rapid growth stage (II). During this period, the large brine pools that facilitate the precipitation of large individual crystals are mostly depleted. Alternatively, the increased growth rate leads to the nucleation and precipitation of tiny aggregated salt accumulations.

    A comparison between the two surface wettability cases suggests that brine continuity and accessibility significantly impact crystallization dynamics and morphology. In the hydrophobic chip, crystal growth was dominated by crystallization at the water-gas interface, where individual crystalline halites formed. Precipitation and growth are closely tied to specific salting-out sites with minimal displacement from the original brine locations. This results in a cleaner precipitation pattern within the pore network of the hydrophobic chip, as shown in Figure~\ref{fig:figure4}b. This morphology suggests potentially less dramatic adverse effects on permeability deterioration, which is crucial when constructing clogging models and defining the porosity-permeability relationship for reactive transport modeling. Developing precise clogging models to accurately represent the impact of porous geometry alterations induced by solid precipitation on fluid flow properties remains a substantial challenge \cite{Masoudi20249988, MASOUDI2021103475, parvin2020continuum}.
    
    For the hydrophilic chip, polycrystalline aggregates begin to appear, forming dendritic or tree-like porous structures due to the continuity and conductivity of the brine films. These features are often characterized by diffusion-limited growth supported by capillary-driven brine suction, as shown in the last two precipitation time frames of Figure~\ref{fig:figure4} (top row). The initial porous, semi-dendritic growth of micrometer-sized halite aggregates increases the evaporation surface area, significantly enhancing capillary-driven brine suction within the porous medium, and promoting salt's self-enhancing growth. At the shrinking aqueous phase interface, capillary forces increase the brine surface curvature, confining the brine and creating ideal conditions for nucleation sites. In these regions, the local supersaturation of dissolved salts significantly lowers the energy barrier for subsequent halite nucleation, thereby promoting enhanced salt crystal formation. The brine films on water-wet surfaces transport ions from other brine patches, enabling more ions to participate in salt aggregation. As the micrometer-sized aggregates, which are comprised of nano-micron-sized crystals, connect and merge, they form larger and more complex structures that gradually fill the pore space \cite{shahidzadeh2015salt, Shokri-Kuehni2017}. This self-enhancing process can lead to substantial salt crystal assemblies within the pores even before complete drying occurs.
	
    During stage III, the crystallization morphology continues to evolve as long as brine is available, predominantly on substrate surfaces that support residual brine movement. The remaining solutes in the chips can migrate toward the precipitated aggregation sites due to strong capillary-induced flow. The hydrophilic nature of salt crystals also helps to retain these delivered ions on the crystal surface, making the final aggregate structure more complex compared to the initial interconnections of micrometer-sized, cubic, or semi-crystalline halite structures. Moreover, at this stage, the porous or semi-porous aggregates could become non-porous or develop sub-resolution intercrystalline porosity (nano-porous structures that cannot be resolved with the conducted imaging).

    \subsection{Spatial distribution of salt crystals}
\begin{figure}
    \centering	
    \includegraphics[width=0.9\textwidth]{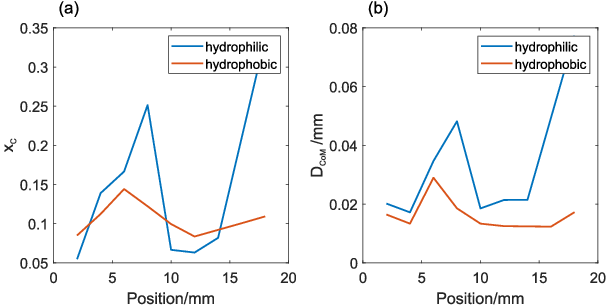}
    \caption{Analysis of the spatial distribution of salt accumulations within hydrophilic and hydrophobic microfluidic chips after the experiments. (a) Crystal areal coverage ($X_c$). (b) Sizes of precipitates, characterized by the equivalent diameter of patches ($D_{\text{CoM}}$).}
    \label{fig:figure6}
    \end{figure}

    Figure~\ref{fig:figure6} illustrates the spatial distribution of salt precipitation along the chips at various locations after the porous structures have completely dried. The wettability of the substrate surfaces influences the areal coverage ($X_c$) and the sizes of precipitates, characterized by the equivalent diameter of patches ($D_{\text{CoM}}$). As described above, hydrophobic porous networks exhibit slower evaporation and salt precipitation rates than water-wet surfaces. Consequently, hydrophilic substrates tend to form larger patches of salt crystals and achieve greater overall areal coverage along the chip, transitioning from the inlet to the outlet.
    
    As Figure~\ref{fig:figure6} indicates, a substantial amount of salt crystals accumulates near the outlet of the hydrophilic chip. This may occur due to the capillary-end effect in the water-wet pores \cite{huang1998capillary, guedon2017influence, naseryan2019relative}, resulting in higher water saturation near the outlet and providing a necessary solute source for salt formation once supersaturation is achieved. Smaller and medium-sized brine pools are more broadly distributed around the area close to the injection channel after most of the brine has been displaced by the gas. In contrast, the hydrophobic chip exhibits a relatively uniform spatial distribution with significantly lower extremes in areal coverage and equivalent sizes throughout.    
    
    The weaker adhesion between water and solid surfaces diminishes the capillary forces responsible for water movement through the pore spaces in hydrophobic pores. The brine continuity and capillarity cannot be sustained because there are no brine films acting as the corner-wetting phase when the contact angle exceeds 90$^\circ$ \cite{kijlstra2002roughness, shokri2008effects}. As evaporation progresses, the hydrophobicity eventually disrupts the initial connectivity of the aqueous phase, thereby reducing brine capillary movements.
    
    The precipitation pattern in the hydrophilic chip, characterized by banded and aggregated growth, creates local extremities that can obstruct flow pathways due to the volume of salt growth within the pore network. This results in a moderate decline in porosity and a significant reduction in permeability, implying that CO$_2$-induced salt precipitation can hinder the accessible pathways for the injected gas and potentially damage the pore structure.
   
    Substantial aggregated salt precipitation occurs during the rapid growth stage near the extended brine pools. Subsequently, the remaining brine is transported and concentrated around the adjacent salt deposits, leading to increased local salt precipitation and expanded coverage. This process aligns with the self-enhancing mechanisms previously proposed and discussed \cite{miri2016review, nooraiepour2018salt}. The hydrophilic nature of the salt plays a crucial role in sustaining a brine film on the surfaces of the salt aggregates. This liquid film facilitates the movement of residual brine pools toward the precipitated salt areas. As water evaporates from the brine film, fresh brine is continuously supplied via the hydrophilic substrate from the pool to the drying surface, thereby aiding the accelerated growth in two localities with a sharp increase in salt accumulations, as shown in Figure~\ref{fig:figure6} for the hydrophilic chip.

    \subsection{Probabilistic nature of salt nucleation and precipitation}
\begin{figure}
    \centering	
    \includegraphics[width=0.9\textwidth]{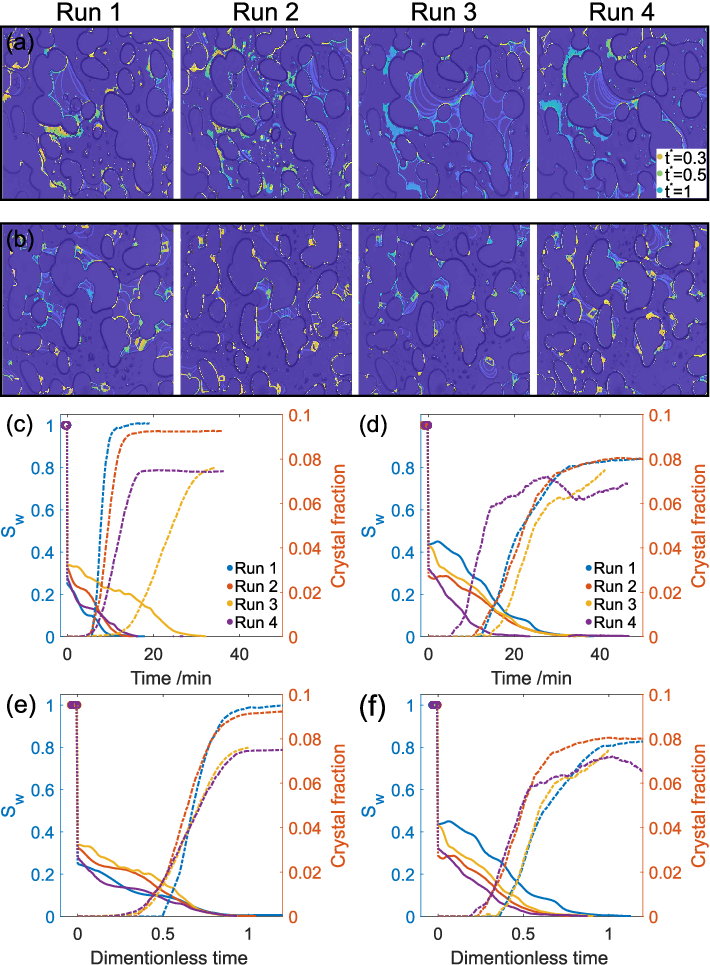}
    \caption{Probabilistic spatial and temporal dynamics of halite nucleation and growth across four repeated experiments for (a) hydrophilic and (b) hydrophobic chips. Each figure presents stacked experimental images capturing four subsequent time steps, as detailed in Figure~\ref{fig:figure3}. Brine saturation reduction (solid lines) and crystal growth (dashed lines) are plotted as functions of elapsed time under a 2 ml/min CO$_2$ flow for (c, d) hydrophilic and (e, f) hydrophobic chips, respectively.}
    \label{fig:figure5}
    \end{figure}

    Each test was repeated four times to document and quantify the probabilistic nature of halite nucleation and crystal growth within the porous medium, as well as the randomness observed during our reactive transport experiments. Figure~\ref{fig:figure5}(a, b) illustrates temporal snapshots of brine saturation reduction and halite growth for both hydrophilic and hydrophobic chips. Each color map consists of a stack of four snapshots taken at four temporal stages: $t^* = 0; 0.3; 0.5; 1$, as shown in Figure~\ref{fig:figure3}, where the temporal dynamics are superimposed onto the matrix grain structure of the chips. The respective numerical values of $S_w$ and $X_c$ are also plotted over experimental time (Figs.~\ref{fig:figure5}c-f). The graphs present stochastic variations in saturation and growth profiles based on the image processing of microfluidic tests. The left axis (in blue) with solid lines indicates brine saturation, while the right axis (in red) with dashed lines represents crystal areal coverage.

    For the conducted measurements, crystallization on the hydrophilic chip begins, on average, after 8 minutes and concludes 17 minutes after the start of CO$_2$ flow. For the hydrophobic chip, crystallization starts at 10 minutes and ends at 28 minutes. This indicates that crystallization begins earlier and progresses faster on the hydrophilic chip, suggesting that surface wettability impacts crystallization kinetics. In both scenarios, crystallization initiates when $S_w$ reaches approximately 50\% of its initial value after drainage. Since nearly saturated brine was used in the experiments, a significant drop in brine saturation before the onset of crystallization suggests that a portion of the brine is displaced out of the chips by the CO$_2$ stream. Otherwise, crystallization would begin with minimal change in $S_w$.

    The color map of repeated experiments demonstrates greater statistical variations in the spatial location and amount of salt precipitation for the hydrophobic chip. Initial residual brine saturation varies between 25-34\% for the hydrophilic chip and 29-44\% for the hydrophobic chip, indicating a scatter of 9\% and 15\%, respectively. This suggests that drainage sweeps more brine out of the hydrophilic chip compared to the hydrophobic one. However, the precipitating areal coverage $X_c$ ranges between 7.5-9.6\% for the hydrophilic chip and 6.9-8.1\% for the hydrophobic chip, showing a variation of 2.1\% versus 1.2\%, respectively. The variations in $X_c$ indicate that although the hydrophilic chip consistently exhibits lower initial residual $S_w$, due to the continuity and conductivity of the wetting phase on the grains, it accumulates more salt.

    Comparing different experimental repetitions reveals variability in the locations of residual brine patches after viscous displacement. Consequently, the sites of salt crystal nucleation and growth differ between experiments. However, certain regions within the chips show consistent crystal growth across multiple runs. The spatial randomness of growth locations is significantly higher for the hydrophobic chip (Fig.~\ref{fig:figure5}b). In contrast, as shown in Figure~\ref{fig:figure5}a for the hydrophilic chip, there are consistent regions of brine film movement that contribute to capillary flow and aggregation, repeating across different experimental runs.

\section[Implications for mitigating CO2-induced salt pore clogging]{Implications for mitigating CO$_2$-induced salt pore clogging}
    Injecting CO$_2$ into deep (hyper)saline aquifers eventually triggers salt precipitation and growth within the porous medium of the reservoir, originating from the in-situ brine saturation in the rocks. This brine is displaced and evaporated due to the continuous injection of supercritical CO$_2$ at a million tons per year scale per injection well. The resulting salt accumulations clog pore throats and pore bodies, reducing porosity and deteriorating permeability. This issue is most pronounced near the injection site due to the so-called wellbore dry-out, posing a substantial threat to the entire sequestration process, particularly from the perspectives of injectivity and containment.

    The decrease in injectivity caused by lowered permeability and excess pressure buildup from growth has several consequences. Increased injection costs arise as more energy is required to maintain injection rates through obstructed pores. Additionally, reduced permeability compromises storage efficiency as the injected CO$_2$ cannot disperse effectively throughout the saline reservoir. This limited injectivity further restricts the aquifer's overall storage potential, undermining the overarching objective of reducing atmospheric carbon emissions on a global scale. Furthermore, pressure buildup can compromise the (geo)mechanical integrity of the reservoir. Triaxial mechanical tests conducted on salt-affected sandstones have demonstrated that halite crystallization substantially weakens the mechanical strength and deteriorates the rock's microstructure at both the pore and grain scales \cite{nooraiepour2024Damage}. The necessary workover and cessation of operations to address and remediate near-wellbore salt accumulation introduce additional economic, operational, and safety hurdles.

    As described, water-wet surfaces and capillary forces in porous media significantly enhance the evaporation process due to their intrinsic properties and synergistic effects. Water-wet surfaces ensure the continuous adherence and spread of thin water films. This adherence prevents rapid dehydration of the porous network and increases the surface area available for evaporation. Moreover, thin film evaporation is more efficient than bulk water evaporation, owing to the larger surface-to-volume ratio. 

    Capillary forces in porous media are pivotal in sustaining and uniformly distributing water, preventing dry spots, and promoting balanced evaporation. Additionally, these forces enhance the molecular mobility of water, thus supporting a consistent evaporation rate. Capillary forces also facilitate capillary-driven backflow towards the evaporation front and promote the secondary nucleation and precipitation of halite crystals.

    Even without accounting for the self-enhancing nature of salt growth, the combined effects of water-wet surfaces and capillary forces create a self-sustaining system that significantly optimizes the evaporation process. The synergy between these factors ensures a sustained evaporation rate, continuously supplying water to the surface and maintaining high evaporation rates even when bulk water is not directly exposed. In other words, water-wet substrate surfaces create a continuous path that maintains hydraulic connectivity to the evaporating surface through capillary action. This combination also helps maintain a near-thermodynamic equilibrium condition at the evaporative interface, which is essential for efficient phase change processes. Additionally, by reducing the overall energy barrier for evaporation, this interplay enhances the energy efficiency of the process.

    The results of the present experimental study suggest that an effective solution to address the issue of CO$_2$-induced salt precipitation could involve modifying the surface properties of the reservoir rocks through wettability alteration. Wettability plays a critical role in the distribution and flow dynamics of fluids within porous media. By manipulating the wettability of the rocks at the near-wellbore area, it is possible not only to mitigate salt growth challenges but also to potentially enhance CO$_2$ injection efficiency.

    \begin{figure}
		\centering	
		\includegraphics[width=0.9\textwidth]{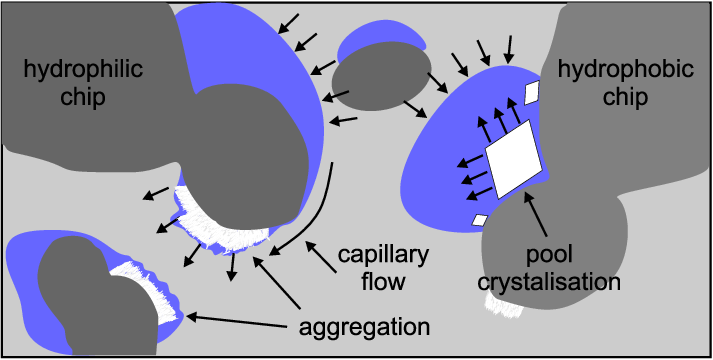}
		\caption{Schematic representation of CO$_2$-induced salt precipitation and growth morphologies in hydrophilic and hydrophobic chips. In hydrophilic chips, the continuity of the brine phase and strong capillary flow facilitates extensive salt aggregation. Conversely, in hydrophobic chips, weaker and localized capillary-driven backflow results in limited salt aggregation.}
		\label{fig:figure7}
    \end{figure}

    Figure~\ref{fig:figure7} presents a schematic illustrating the two primary halite crystallization morphologies observed in our experiments. The surface wettability of the chips controls brine distribution and solute transport during different stages of the experiments. In hydrophilic networks, brine preferentially adheres to the substrate surfaces, forming thin films and potentially corner flows that enhance connectivity throughout the porous medium. This leads to a distributed brine network and facilitates the transport of dissolved solutes across a larger surface area on the chip. Consequently, these conditions promote more widespread and potentially faster rates of crystallization during evaporation, supported by continuous solute transport as the solutes encounter numerous nucleation sites.

    In contrast, on hydrophobic surfaces, residual brine tends to avoid the grain surfaces and instead forms isolated clusters or droplets within the pore spaces. This results in less connected water pathways and more limited solute transport, which can hinder the continuous movement of solutes and may slow down the overall crystallization process. While the reduced interaction between the solutes and the pore walls in hydrophobic media can lead to fewer nucleation events, the growth dynamics are limited due to restricted access to brine pools.   	

    As a result, when the substrate surface becomes less hydrophilic, the brine tends to be more uniformly distributed, preventing localized evaporation and subsequent salt crystal formation in pore throats. This results in less pore blockage, preserving the porosity and permeability. Effective wettability alteration ensures that the injected CO$_2$ can move more freely through the pore spaces, minimizing the risk of wellbore dry-out and associated injectivity issues.

    With improved wettability conditions, the CO$_2$ and brine can coexist more stably within the porous network. This stabilized dispersion helps maintain higher injectivity rates, reduces the energy and cost requirements for injection operations, and enhances overall carbon sequestration efficiency. Moreover, it has been reported that an intermediate-wet condition often leads to a more favorable capillary trapping mechanism, which can help securely store CO$_2$ in the subsurface \cite{hu2017wettability, alyafei2016effect}.

    \section{Conclusions}
    \label{Conclusions}
    The study revealed distinct CO$_2$-brine interface configurations influenced by surface wettability. In hydrophilic chips, larger brine clusters and fewer flow channels indicated enhanced wetting phase connectivity. We identified three stages of CO$_2$-induced brine evaporation and salt precipitation, with a notable rapid growth phase during stage II due to fast halite nucleation and growth rates, along with sufficient gas-liquid contact areas for evaporation.
    
    Two primary halite morphologies were observed: large individual crystals and polycrystalline aggregates of micrometer-sized halite. These aggregates formed intercrystalline porous structures within the original pore geometries. Comparisons between surface wettability demonstrated that brine continuity and accessibility significantly influenced crystallization dynamics and morphology. In hydrophobic chips, crystal growth primarily occurred in residual brine pools, leading to cleaner precipitation patterns and minor permeability deterioration. Conversely, hydrophilic chips exhibited dendritic polycrystalline aggregates that gradually filled pore spaces.
    
    Hydrophilic chips showed more extreme and varied halite accumulations, while hydrophobic chips displayed a relatively uniform spatial distribution with lower extremes in areal coverage and crystal sizes. In hydrophobic pores, weaker adhesion decreases capillary forces, disrupting brine continuity and reducing capillary movements. The precipitation pattern in hydrophilic chip, characterized by banded and aggregated growth, could obstruct flow pathways, leading to moderate porosity decline and significant permeability reduction.
    
    Experimental repetitions revealed probabilistic spatial variability in residual brine patches, leading to differing nucleation and growth sites. The probabilistic range was higher in hydrophilic chip, yet these water-wet surfaces consistently showed regions of brine film movement contributing to aggregation across multiple runs.
    
    The combined effects of water-wet surfaces and capillary forces create a self-sustaining system that optimizes evaporation by maintaining hydraulic connectivity and high evaporation rates. These factors help maintain near-thermodynamic equilibrium at the evaporative interface, enhancing energy efficiency. Based on the research outcomes, we suggest that modifying reservoir rock surface properties near the injection well through wettability alteration could effectively address salt precipitation issues.

    \section*{Data availability}
    The data supporting the findings of this study are publicly available in the RODBUK open repository via DOI: \href{https://doi.org/10.58032/AGH/TRKY5F}{10.58032/AGH/TRKY5F}.
						
    \section*{Acknowledgments}						
    This work was supported by the “solid and salt precipitation kinetics during CO$_2$ injection into reservoir” project, funded by Norway Grants (Norwegian Financial Mechanism 2014-2021) under grant number UMO-2019/34/H/ST10/00564. This research was also partially funded by EEA and Norway Grants under grant number NOR/POLNORCCS/AGaStor/0008/2019-00.

    \bibliography{mybibfile}

\newcommand{\noop}[1]{}
\begin{thebibliography}{10}

\bibitem{IPCC_2022_WGIII}
IPCC.
\newblock {\em Climate Change 2022: Mitigation of Climate Change. Contribution of Working Group III to the Sixth Assessment Report of the Intergovernmental Panel on Climate Change}.
\newblock Cambridge University Press, Cambridge, UK and New York, NY, USA, 2022.

\bibitem{ringrose2021storage}
Philip~S Ringrose, Anne-Kari Furre, Stuart~MV Gilfillan, Samuel Krevor, Martin Landr{\o}, Rory Leslie, Tip Meckel, Bamshad Nazarian, and Adeel Zahid.
\newblock Storage of carbon dioxide in saline aquifers: physicochemical processes, key constraints, and scale-up potential.
\newblock {\em Annual Review of Chemical and Biomolecular Engineering}, 12(1):471--494, 2021.

\bibitem{oelkers2023carbon}
Eric~H Oelkers and Sigurdur~R Gislason.
\newblock Carbon capture and storage: From global cycles to global solutions.
\newblock {\em Geochemical Perspectives}, 12(2):179--180, 2023.

\bibitem{gislason2014carbon}
Sigurdur~R Gislason and Eric~H Oelkers.
\newblock Carbon storage in basalt.
\newblock {\em Science}, 344(6182):373--374, 2014.

\bibitem{sandalow2021carbon}
D~Sandalow, R~Aines, J~Friedmann, P~Kelemen, C~McCormick, I~Power, B~Schmidt, and S~Wilson.
\newblock Carbon mineralization roadmap draft october 2021.
\newblock Technical report, Lawrence Livermore National Lab.(LLNL), Livermore, CA (United States), 2021.

\bibitem{krevor2023subsurface}
Samuel Krevor, Heleen De~Coninck, Sarah~E Gasda, Navraj~Singh Ghaleigh, Vincent de~Gooyert, Hadi Hajibeygi, Ruben Juanes, Jerome Neufeld, Jennifer~J Roberts, and Floris Swennenhuis.
\newblock Subsurface carbon dioxide and hydrogen storage for a sustainable energy future.
\newblock {\em Nature Reviews Earth \& Environment}, 4(2):102--118, 2023.

\bibitem{Hellevang2015}
Helge Hellevang.
\newblock {\em Carbon Capture and Storage (CCS)}, pages 591--602.
\newblock Springer Berlin Heidelberg, Berlin, Heidelberg, 2015.

\bibitem{nooraiepour2024Damage}
Mohammad Nooraiepour, Krzysztof Polanski, Mohammad Masoudi, Szymon Kuczynski, Hannelore Derluyn, Liebert~Parreiras Nogueira, Bahman Bohloli, Stanislaw Nagy, and Helge Hellevang.
\newblock Assessing salt precipitation and weak acid interaction in subsurface {CO}$_2$ injection: Potential 50\% strength decline in near-wellbore reservoir sandstones.
\newblock {\em Rock Mechanics and Rock Engineering, under arXiv preprint arXiv:2401.14852}, 2024.

\bibitem{shahidzadeh2024crystallization}
Noushine SHAHIDZADEH.
\newblock Crystallization pressure.
\newblock {\em Salt Crystallization in Porous Media}, pages 25--44, 2024.

\bibitem{ringrose2009plume}
Philip Ringrose, Mansour Atbi, David Mason, Marianne Espinassous, {\O}yvind Myhrer, Martin Iding, Allan Mathieson, and Iain Wright.
\newblock Plume development around well kb-502 at the in salah {CO}$_2$ storage site.
\newblock {\em First break}, 27(1), 2009.

\bibitem{zettlitzer2010re}
Michael Zettlitzer, Fabian Moeller, Daria Morozova, Peter Lokay, Hilke W{\"u}rdemann, {CO2SINK} Group, et~al.
\newblock Re-establishment of the proper injectivity of the {CO}$_2$-injection well ktzi 201 in ketzin, germany.
\newblock {\em International Journal of Greenhouse Gas Control}, 4(6):952--959, 2010.

\bibitem{Khosravi2023}
{\em {Challenges in Simulation of Salt Clogging}}, volume SPE EuropEC - Europe Energy Conference featured at the 84th EAGE Annual Conference \& Exhibition of {\em SPE Europec featured at EAGE Conference and Exhibition}, 06 2023.

\bibitem{torsaeter2018geological}
Malin Tors{\ae}ter and Pierre Cerasi.
\newblock Geological and geomechanical factors impacting loss of near-well permeability during {CO}$_2$ injection.
\newblock {\em International Journal of Greenhouse Gas Control}, 76:193--199, 2018.

\bibitem{masoudi2021thesis}
Mohammad Masoudi.
\newblock {\em Near wellbore processes during carbon capture, utilization, and storage (CCUS): an integrated modeling approach}.
\newblock PhD thesis, University of Oslo, 2021.
\newblock Available at \url{http://urn.nb.no/URN:NBN:no-92465}.

\bibitem{XIAO2024ReviewRisk}
Ting Xiao, Ting Chen, Zhiwei Ma, Hailong Tian, Saro Meguerdijian, Bailian Chen, Rajesh Pawar, Lianjie Huang, Tianfu Xu, Martha Cather, and Brian McPherson.
\newblock A review of risk and uncertainty assessment for geologic carbon storage.
\newblock {\em Renewable and Sustainable Energy Reviews}, 189:113945, 2024.

\bibitem{nooraiepour2018effect}
Mohammad Nooraiepour, Hossein Fazeli, Rohaldin Miri, and Helge Hellevang.
\newblock Effect of {CO}$_2$ phase states and flow rate on salt precipitation in shale caprocks—a microfluidic study.
\newblock {\em Environmental science \& technology}, 52(10):6050--6060, 2018.

\bibitem{cui2023review}
Guodong Cui, Zhe Hu, Fulong Ning, Shu Jiang, and Rui Wang.
\newblock A review of salt precipitation during {CO}$_2$ injection into saline aquifers and its potential impact on carbon sequestration projects in china.
\newblock {\em Fuel}, 334:126615, 2023.

\bibitem{miri2016review}
Rohaldin Miri and Helge Hellevang.
\newblock Salt precipitation during {CO}$_2$ storage—a review.
\newblock {\em International Journal of Greenhouse Gas Control}, 51:136--147, 2016.

\bibitem{derluyn2024experimental}
Hannelore DERLUYN.
\newblock Experimental observations on salt crystallization in geomaterials.
\newblock {\em Salt Crystallization in Porous Media}, pages 99--125, 2024.

\bibitem{nooraiepour2018salt}
Mohammad Nooraiepour, Hossein Fazeli, Rohaldin Miri, and Helge Hellevang.
\newblock Salt precipitation during injection of {CO}$_2$ into saline aquifers: lab-on-chip experiments on glass and geomaterial microfluidic specimens.
\newblock In {\em 14th Greenhouse Gas Control Technologies Conference Melbourne}, pages 21--26, 2018.

\bibitem{talman2020salt}
Stephen Talman, Alireza~Rangriz Shokri, Rick Chalaturnyk, and Erik Nickel.
\newblock Salt precipitation at an active {CO}$_2$ injection site.
\newblock {\em Gas Injection into Geological Formations and Related Topics}, pages 183--199, 2020.

\bibitem{sokama2023experimental}
Yen~Adams Sokama-Neuyam, Muhammad Aslam~Md Yusof, Shadrack~Kofi Owusu, Victor Darkwah-Owusu, Joshua~Nsiah Turkson, Adwoa~Sampongmaa Otchere, and Jann~Rune Ursin.
\newblock Experimental and theoretical investigation of the mechanisms of drying during {CO}$_2$ injection into saline reservoirs.
\newblock {\em Scientific reports}, 13(1):9155, 2023.

\bibitem{mim2023minireview}
Rubaya~Tasnin Mim, Berihun~Mamo Negash, Shiferaw~Regassa Jufar, and Faizan Ali.
\newblock Minireview on {CO}$_2$ storage in deep saline aquifers: Methods, opportunities, challenges, and perspectives.
\newblock {\em Energy \& Fuels}, 37(23):18467--18484, 2023.

\bibitem{MIRI201510}
Rohaldin Miri, Reinier {van Noort}, Per Aagaard, and Helge Hellevang.
\newblock New insights on the physics of salt precipitation during injection of {CO}$_2$ into saline aquifers.
\newblock {\em International Journal of Greenhouse Gas Control}, 43:10--21, 2015.

\bibitem{ott2015salt}
H~Ott, SM~Roels, and K~De~Kloe.
\newblock Salt precipitation due to supercritical gas injection: I. capillary-driven flow in unimodal sandstone.
\newblock {\em International Journal of Greenhouse Gas Control}, 43:247--255, 2015.

\bibitem{ott2021salt}
Holger Ott, Jeroen Snippe, and Kees de~Kloe.
\newblock Salt precipitation due to supercritical gas injection: Ii. capillary transport in multi porosity rocks.
\newblock {\em International Journal of Greenhouse Gas Control}, 105:103233, 2021.

\bibitem{NOROUZI2021CO2Salt}
Amir~Mohammad Norouzi, Masoud Babaei, Weon~Shik Han, Kue-Young Kim, and Vahid Niasar.
\newblock Co2-plume geothermal processes: A parametric study of salt precipitation influenced by capillary-driven backflow.
\newblock {\em Chemical Engineering Journal}, 425:130031, 2021.

\bibitem{MASOUDI2021103475}
Mohammad Masoudi, Hossein Fazeli, Rohaldin Miri, and Helge Hellevang.
\newblock Pore scale modeling and evaluation of clogging behavior of salt crystal aggregates in {CO}$_2$-rich phase during carbon storage.
\newblock {\em International Journal of Greenhouse Gas Control}, 111:103475, 2021.

\bibitem{HoTsai2020}
Tsai-Hsing~Martin Ho and Peichun~Amy Tsai.
\newblock Microfluidic salt precipitation: implications for geological co2 storage.
\newblock {\em Lab Chip}, 20:3806--3814, 2020.

\bibitem{Falcon-Suarez2020}
Ismael~Himar Falcon-Suarez, Kurt Livo, Ben Callow, Hector Marin-Moreno, Manika Prasad, and Angus~Ian Best.
\newblock Geophysical early warning of salt precipitation during geological carbon sequestration.
\newblock {\em Scientific Reports}, 10(1), 2020.
\newblock Cited by: 14; All Open Access, Gold Open Access.

\bibitem{LIMA2021FracDry}
Marina~Grimm Lima, Hoda Javanmard, Daniel Vogler, Martin~O. Saar, and Xiang-Zhao Kong.
\newblock Flow-through drying during co2 injection into brine-filled natural fractures: A tale of effective normal stress.
\newblock {\em International Journal of Greenhouse Gas Control}, 109:103378, 2021.

\bibitem{ZEIDOUNI2009600}
Mehdi Zeidouni, Mehran Pooladi-Darvish, and David Keith.
\newblock Analytical solution to evaluate salt precipitation during co2 injection in saline aquifers.
\newblock {\em International Journal of Greenhouse Gas Control}, 3(5):600--611, 2009.

\bibitem{Shokri2024soilSal}
Nima Shokri, Amirhossein Hassani, and Muhammad Sahimi.
\newblock Multi-scale soil salinization dynamics from global to pore scale: A review.
\newblock {\em Reviews of Geophysics}, 62(4):e2023RG000804, 2024.
\newblock e2023RG000804 2023RG000804.

\bibitem{Datta2023LabCarbon}
Sujit~S. Datta, Ilenia Battiato, Martin~A. Fernø, Ruben Juanes, Shima Parsa, Valentina Prigiobbe, Enric Santanach-Carreras, Wen Song, Sibani~Lisa Biswal, and David Sinton.
\newblock Lab on a chip for a low-carbon future.
\newblock {\em Lab on a Chip}, 23(5):1358 – 1375, 2023.
\newblock Cited by: 12; All Open Access, Bronze Open Access.

\bibitem{Morai2020HPMicrofluidic}
Sandy Morais, Anaïs Cario, Na~Liu, Dominique Bernard, Carole Lecoutre, Yves Garrabos, Anthony Ranchou-Peyruse, Sébastien Dupraz, Mohamed Azaroual, Ryan~L. Hartman, and Samuel Marre.
\newblock Studying key processes related to co2 underground storage at the pore scale using high pressure micromodels.
\newblock {\em React. Chem. Eng.}, 5:1156--1185, 2020.

\bibitem{HE2022ExpSalt}
Di~He, Ruina Xu, Tiancheng Ji, and Peixue Jiang.
\newblock Experimental investigation of the mechanism of salt precipitation in the fracture during co2 geological sequestration.
\newblock {\em International Journal of Greenhouse Gas Control}, 118:103693, 2022.

\bibitem{dTzachMineralWett}
Andreas Tzachristas, Roxanne-Irene Malamoudis, Dimitra~G. Kanellopoulou, Eugene Skouras, John Parthenios, Petros~G. Koutsoukos, Christakis~A. Paraskeva, and Varvara Sygouni.
\newblock Mineral scaling in microchips: Effect of substrate wettability on caco3 precipitation.
\newblock {\em Industrial \& Engineering Chemistry Research}, 59(45):20201--20210, 2020.

\bibitem{DARKWAHOWUSU2024Review}
A comprehensive review of remediation strategies for mitigating salt precipitation and enhancing co2 injectivity during co2 injection into saline aquifers.
\newblock {\em Science of The Total Environment}, 950:175232, 2024.

\bibitem{akindipe2022salt}
Dayo Akindipe, Soheil Saraji, and Mohammad Piri.
\newblock Salt precipitation in carbonates during supercritical {CO}$_2$ injection: A pore-scale experimental investigation of the effects of wettability and heterogeneity.
\newblock {\em International Journal of Greenhouse Gas Control}, 121:103790, 2022.

\bibitem{zhang2024brine}
Hui Zhang, Zhonghao Sun, Nan Zhang, and Budi Zhao.
\newblock Brine drying and salt precipitation in porous media: A microfluidics study.
\newblock {\em Water Resources Research}, 60(1):e2023WR035670, 2024.

\bibitem{yan2024dynamics}
Lifei Yan, Rustam Niftaliyev, Denis Voskov, and Rouhi Farajzadeh.
\newblock Dynamics of salt precipitation at pore scale during {CO}$_2$ subsurface storage in saline aquifer.
\newblock {\em Journal of Colloid and Interface Science}, 2024.

\bibitem{he2019pore}
Di~He, Peixue Jiang, and Ruina Xu.
\newblock Pore-scale experimental investigation of the effect of supercritical {CO}$_2$ injection rate and surface wettability on salt precipitation.
\newblock {\em Environmental Science \& Technology}, 53(24):14744--14751, 2019.

\bibitem{rufai2018effect}
Ayorinde Rufai and John Crawshaw.
\newblock Effect of wettability changes on evaporation rate and the permeability impairment due to salt deposition.
\newblock {\em ACS Earth and Space Chemistry}, 2(4):320--329, 2018.

\bibitem{shahidzadeh2004effect}
N~Shahidzadeh-Bonn, A~Tourni{\'e}, S~Bichon, P~Vi{\'e}, S~Rodts, P~Faure, F~Bertrand, and A~Azouni.
\newblock Effect of wetting on the dynamics of drainage in porous media.
\newblock {\em Transport in porous media}, 56(2):209--224, 2004.

\bibitem{bachmann2001isothermal}
J~Bachmann, R~Horton, and RR~Van Der~Ploeg.
\newblock Isothermal and nonisothermal evaporation from four sandy soils of different water repellency.
\newblock {\em Soil Science Society of America Journal}, 65(6):1599--1607, 2001.

\bibitem{shahidzadeh2007effect}
N~Shahidzadeh-Bonn, A~Azouni, and P~Coussot.
\newblock Effect of wetting properties on the kinetics of drying of porous media.
\newblock {\em Journal of physics: condensed matter}, 19(11):112101, 2007.

\bibitem{SHOKRI2008Hydrophob}
N.~Shokri, P.~Lehmann, and D.~Or.
\newblock Effects of hydrophobic layers on evaporation from porous media.
\newblock {\em Geophysical Research Letters}, 35(19), 2008.

\bibitem{bergstad2016evaporation}
Mina Bergstad and Nima Shokri.
\newblock Evaporation of nacl solution from porous media with mixed wettability.
\newblock {\em Geophysical research letters}, 43(9):4426--4432, 2016.

\bibitem{SHOKRI2013135}
N.~Shokri and D.~Or.
\newblock Drying patterns of porous media containing wettability contrasts.
\newblock {\em Journal of Colloid and Interface Science}, 391:135--141, 2013.

\bibitem{Desarnaud2015salt}
Julie Desarnaud, Hannelore Derluyn, Luisa Molari, Stefano de~Miranda, Veerle Cnudde, and Noushine Shahidzadeh.
\newblock {Drying of salt contaminated porous media: Effect of primary and secondary nucleation}.
\newblock {\em Journal of Applied Physics}, 118(11):114901, 09 2015.

\bibitem{PRAT2024}
Marc PRAT.
\newblock {\em Evaporation, Transport, and Crystallization}, chapter~3, pages 45--74.
\newblock John Wiley \& Sons, Ltd, 2024.

\bibitem{Nooraiepour2021SciRep}
Mohammad Nooraiepour, Mohammad Masoudi, and Helge Hellevang.
\newblock Probabilistic nucleation governs time, amount, and location of mineral precipitation and geometry evolution in the porous medium.
\newblock {\em Scientific Reports}, 11(1), 2021.

\bibitem{Masoudi20249988}
Mohammad Masoudi, Mohammad Nooraiepour, Hang Deng, and Helge Hellevang.
\newblock Mineral precipitation and geometry alteration in porous structures: How to upscale variations in permeability-porosity relationship?
\newblock {\em Energy and Fuels}, 38(11):9988 – 10001, 2024.

\bibitem{Nooraiepour2021Omega}
Mohammad Nooraiepour, Mohammad Masoudi, Nima Shokri, and Helge Hellevang.
\newblock Probabilistic nucleation and crystal growth in porous medium: New insights from calcium carbonate precipitation on primary and secondary substrates.
\newblock {\em ACS Omega}, 6(42):28072 – 28083, 2021.

\bibitem{nooraiepour2022GHGT16}
Mohammad Nooraiepour, Mohammad Masoudi, Nima Shorki, and Helge Hellevang.
\newblock Precipitation-induced geometry evolution in porous media: Numerical and experimental insights based on new model on probabilistic nucleation and mineral growth.
\newblock In {\em Proceedings of the 16th Greenhouse Gas Control Technologies Conference (GHGT-16)}, pages 23--24, 2022.

\bibitem{Fazeli2020nucleation}
Hossein Fazeli, Mohammad Masoudi, Ravi~A. Patel, Per Aagaard, and Helge Hellevang.
\newblock Pore-scale modeling of nucleation and growth in porous media.
\newblock {\em ACS Earth and Space Chemistry}, 4(2):249--260, 2020.

\bibitem{DENG2022105445}
Hang Deng, Mehdi Gharasoo, Liwei Zhang, Zhenxue Dai, Alireza Hajizadeh, Catherine~A. Peters, Cyprien Soulaine, Martin Thullner, and Philippe {Van Cappellen}.
\newblock A perspective on applied geochemistry in porous media: Reactive transport modeling of geochemical dynamics and the interplay with flow phenomena and physical alteration.
\newblock {\em Applied Geochemistry}, 146:105445, 2022.

\bibitem{Hellevang2019Basalt}
{Hellevang, Helge}, {Wolff-Boenisch, Domenik}, and {Nooraiepour, Mohammad}.
\newblock Kinetic control on the distribution of secondary precipitates during co2-basalt interactions.
\newblock {\em 16th International Symposium on Water-Rock Interaction (WRI-16), E3S Web Conf.}, 98:04006, 2019.

\bibitem{EAGE2024Basalt}
M.~Nooraiepour, M.~Masoudi, B.G. Haile, and H.~Hellevang.
\newblock Intricacies of co2-basalt interactions, reactive flow and carbon mineralization: Bridging numerical forecasts to empirical realities.
\newblock {\em 85th EAGE Annual Conference \& Exhibition}, pages 1--5, 2024.

\bibitem{parvin2020continuum}
Saeed Parvin, Mohammad Masoudi, Anja Sundal, and Rohaldin Miri.
\newblock Continuum scale modelling of salt precipitation in the context of {CO}$_2$ storage in saline aquifers with mrst compositional.
\newblock {\em International Journal of Greenhouse Gas Control}, 99:103075, 2020.

\bibitem{lehmann2008characteristic}
Peter Lehmann, Shmuel Assouline, and Dani Or.
\newblock Characteristic lengths affecting evaporative drying of porous media.
\newblock {\em Physical Review E—Statistical, Nonlinear, and Soft Matter Physics}, 77(5):056309, 2008.

\bibitem{or2013advances}
Dani Or, Peter Lehmann, Ebrahim Shahraeeni, and Nima Shokri.
\newblock Advances in soil evaporation physics—a review.
\newblock {\em Vadose Zone Journal}, 12(4):1--16, 2013.

\bibitem{bacchin2021microfluidic}
Patrice Bacchin, Jacques Leng, and Jean-Baptiste Salmon.
\newblock Microfluidic evaporation, pervaporation, and osmosis: from passive pumping to solute concentration.
\newblock {\em Chemical Reviews}, 122(7):6938--6985, 2021.

\bibitem{MOSAVAT20145408}
Nader Mosavat and Farshid Torabi.
\newblock Application of {CO}$_2$-saturated water flooding as a prospective safe {CO}$_2$ storage strategy.
\newblock {\em Energy Procedia}, 63:5408--5419, 2014.
\newblock 12th International Conference on Greenhouse Gas Control Technologies, GHGT-12.

\bibitem{Nooraiepour2023PorMedAlb}
Mohammad Nooraiepour, Mohammad Masoudi, Nima Shokri, and Helge Hellevang.
\newblock {\em Nucleation and crystal growth on the secondary substrate in Album of Porous Media: Structure and Dynamics}.
\newblock 2023.

\bibitem{desarnaud2015drying}
Julie Desarnaud, Hannelore Derluyn, Luisa Molari, Stefano de~Miranda, Veerle Cnudde, and Noushine Shahidzadeh.
\newblock Drying of salt contaminated porous media: Effect of primary and secondary nucleation.
\newblock {\em Journal of Applied Physics}, 118(11), 2015.

\bibitem{masoudi2024EAGE}
M~Masoudi, M~Nooraiepour, and H~Hellevang.
\newblock Understanding the impact of access to continuous brine source on salt precipitation during geological {CO}$_2$ storage.
\newblock In {\em 85th EAGE Annual Conference \& Exhibition (including the Workshop Programme)}, volume 2024, pages 1--5. European Association of Geoscientists \& Engineers, 2024.

\bibitem{nooraiepour2023EAGE}
Mohammad Nooraiepour, Mohammad Masoudi, Helge Hellevang, Karol D{\k{a}}browski, Szymon Kuczy{\'n}ski, Micha{\l} Zaj{\k{a}}c, Stanis{\l}aw Nagy, and Rafa{\l} Smulski.
\newblock Why is it critical to revisit significance and consequences of salt precipitation during {CO}$_2$ injection?
\newblock 2023.

\bibitem{Masoudi2022EAGE}
M.~Masoudi, M.~Nooraiepour, and H.~Hellevang.
\newblock The effect of preferential nucleation sites on the distribution of secondary mineral precipitates.
\newblock volume~3, page 1659 – 1663, 2022.

\bibitem{shahidzadeh2015salt}
Noushine Shahidzadeh, Marthe~FL Schut, Julie Desarnaud, Marc Prat, and Daniel Bonn.
\newblock Salt stains from evaporating droplets.
\newblock {\em Scientific reports}, 5(1):10335, 2015.

\bibitem{Shokri-Kuehni2017}
Salomé M.~S. Shokri-Kuehni, Thomas Vetter, Colin Webb, and Nima Shokri.
\newblock New insights into saline water evaporation from porous media: Complex interaction between evaporation rates, precipitation, and surface temperature.
\newblock {\em Geophysical Research Letters}, 44(11):5504--5510, 2017.

\bibitem{huang1998capillary}
David~D Huang and Matt~M Honarpour.
\newblock Capillary end effects in coreflood calculations.
\newblock {\em Journal of Petroleum Science and Engineering}, 19(1-2):103--117, 1998.

\bibitem{guedon2017influence}
Ga{\"e}l~Raymond Gu{\'e}don, Jeffrey~De’Haven Hyman, Fabio Inzoli, Monica Riva, and Alberto Guadagnini.
\newblock Influence of capillary end effects on steady-state relative permeability estimates from direct pore-scale simulations.
\newblock {\em Physics of Fluids}, 29(12), 2017.

\bibitem{naseryan2019relative}
Javad Naseryan~Moghadam, Mohammad Nooraiepour, Helge Hellevang, Nazmul~Haque Mondol, and Aagaard Per.
\newblock Relative permeability and residual gaseous co2 saturation in the jurassic brentskardhaugen bed sandstones, wilhelm{\o}ya subgroup, western central spitsbergen, svalbard.
\newblock {\em Norwegian Journal of Geology / Norsk Geologisk Tidsskrift}, 99(2):317--328, 2019.

\bibitem{kijlstra2002roughness}
J~Kijlstra, K~Reihs, and A~Klamt.
\newblock Roughness and topology of ultra-hydrophobic surfaces.
\newblock {\em Colloids and Surfaces A: Physicochemical and Engineering Aspects}, 206(1-3):521--529, 2002.

\bibitem{shokri2008effects}
Nima Shokri, Peter Lehmann, and Dani Or.
\newblock Effects of hydrophobic layers on evaporation from porous media.
\newblock {\em Geophysical Research Letters}, 35(19), 2008.

\bibitem{hu2017wettability}
Ran Hu, Jiamin Wan, Yongman Kim, and Tetsu~K Tokunaga.
\newblock Wettability impact on supercritical co2 capillary trapping: Pore-scale visualization and quantification.
\newblock {\em Water Resources Research}, 53(8):6377--6394, 2017.

\bibitem{alyafei2016effect}
Nayef Alyafei and Martin~J Blunt.
\newblock The effect of wettability on capillary trapping in carbonates.
\newblock {\em Advances in Water Resources}, 90:36--50, 2016.

\end{thebibliography}
		%	\section*{References}

    \end{document}